\documentclass[a4paper]{jpconf}
\usepackage{graphicx}
\newcommand{\be}{\begin{equation}}
\newcommand{\ee}{\end{equation}}
\newcommand{\bea}{\begin{eqnarray}}
\newcommand{\eea}{\end{eqnarray}}

\newcommand{\bfk}{\mbox{\boldmath $k$}}

\def\kt{k_\perp}
\newcommand{\bfp}{\mbox{\boldmath $p$}}

\newcommand{\bfP}{\mbox{\boldmath $P$}}
\newcommand{\bfS}{\mbox{\boldmath $S$}}

\newcommand{\pup}{p^\uparrow}

\def\lsim{\mathrel{\rlap{\lower4pt\hbox{\hskip1pt$\sim$}}\raise1pt\hbox{$<$}}}
\def\gsim{\mathrel{\rlap{\lower4pt\hbox{\hskip1pt$\sim$}}\raise1pt\hbox{$>$}}}
\def\nostrocostruttino#1\over#2{\mathrel{\mathop{\kern 0pt \rlap
{\hbox{$#1$}}} \hbox{\kern-.135em $#2$}}}

\begin{document}

\title{New insight on the Sivers transverse momentum dependent distribution function\footnote{Talk delivered by M Boglione at SPIN2010, 
September 27-October 2, 2010, J\"ulich, Germany.}}

\author{M Anselmino$^{1,2}$, M Boglione$^{1,2}$, U D'Alesio$^{3,4}$, \\
S Melis$^{2,5}$, F Murgia$^4$ and A Prokudin$^6$}

\address{$^1$ Dipartimento di Fisica Teorica, Universit\`a di Torino,
             Via P.~Giuria 1, I-10125 Torino, Italy}
\address{$^2$ INFN, Sezione di Torino, Via P.~Giuria 1, I-10125 Torino, Italy}
\address{$^3$ Dipartimento di Fisica, Universit\`a di Cagliari, I-09042 Monserrato (CA), Italy}
\address{$^4$ INFN, Sezione di Cagliari, C.P.~170, I-09042 Monserrato (CA), Italy}
\address{$^5$ Dipartimento di Scienze e Tecnologie Avanzate, Universit\`a del Piemonte Orientale, \\
             Viale T.~Michel 11, I-15121 Alessandria, Italy}
\address{$^6$ Jefferson Laboratory, 12000 Jefferson Avenue, Newport News, VA 23606}

\ead{boglione@to.infn.it}

\begin{abstract}

Polarised Semi-Inclusive Deep Inelastic Scattering (SIDIS) processes,
$\ell(S_\ell) \, p(S) \to \ell \, h \, X$ allow to study Transverse Momentum
Dependent partonic distributions (TMDs), which reveal a non trivial three 
dimensional internal structure of the hadrons in momentum space. One of the 
most representative of the TMDs is the so-called Sivers function that 
describes the distribution of unpolarized quarks inside a transversely 
polarized proton. We present a novel extraction of the Sivers distribution functions from the most recent experimental data of HERMES and COMPASS 
experiments. Using suitable parametrizations, within the TMD factorization 
scheme, and a simple fitting strategy, we also perform a preliminary 
exploration of the role of the proton sea quarks.  
\end{abstract}


The Sivers distribution function $\Delta^N \! f_ {q/\pup}(x,\kt)$ embodies 
the correlation between the nucleon spin and the quark intrinsic transverse 
momentum and gives the number density of unpolarized quarks $q$ with 
intrinsic transverse momentum $\bfk _\perp$ inside a transversely polarized 
proton $\pup$, with three-momentum $\bfP$ and spin polarization vector 
$\bfS$,
\bea
\hat f_ {q/\pup} (x,\bfk_\perp) &=& f_ {q/p} (x,\kt) +
\frac{1}{2} \, \Delta^N \! f_ {q/\pup}(x,\kt)  \;
{\bfS} \cdot (\hat {\bfP}  \times
\hat{\bfk}_\perp) \,.\label{sivnoi}
\eea
$f_ {q/p}(x,\kt)$ is the unpolarized $x$ and $\kt$ dependent parton 
distribution, and the mixed product ${\bfS} \cdot (\hat {\bfP}  \times \hat{\bfk}_\perp)$ gives the parity invariant azimuthal dependence.
The distribution of quarks in polarized hadrons is not axially symmetric, 
and shows a non trivial 3-D structure. The conceptual importance of the 
Sivers function, its relation with the color QCD dynamics and its gauge 
properties, are being more and more understood. Dedicated experiments
are running, under construction or planned.     

In Ref.~\cite{Anselmino:2008sga}, we presented an extraction of the Sivers 
distribution functions based on a fit of SIDIS experimental data from the 
HERMES and COMPASS collaborations. At that time, the only direct evidence 
of a non-zero Sivers effect came from HERMES experimental data on the 
azimuthal moment $A_{UT}^{\sin(\phi_h-\phi_S)}$, measured for pion and kaon 
production off a proton target~\cite{Diefenthaler:2007rj}; the COMPASS 
collaboration, instead, found a Sivers effect compatible with zero for 
SIDIS off a deuteron target~\cite{:2008dn}. The extraction of the Sivers 
functions was made even more challenging by the fact that the 
$A_{UT}^{\sin(\phi_h-\phi_S)}$ measured by HERMES for $K^+$ was surprisingly 
large, about twice as much as the analogous asymmetry for $\pi^+$ production, 
which hinted at a possible important role of the sea 
quarks~\cite{Anselmino:2008sga}.

Since then, new experimental results have become available: COMPASS have 
released new data on SIDIS production of spinless hadrons off a proton 
target, showing a clear Sivers asymmetry~\cite{Alekseev:2010rw}; a new 
HERMES data analysis, based on a much larger statistics, while confirming 
the previous pion data, softens the enhanced peak in the $K^+$ Sivers 
azimuthal moment~\cite{:2009ti}. 

It is then timely and useful to perform a new fit, trying to understand 
the most essential features of the Sivers distribution function 
as shown by the present (scarce) experimental data.

The SIDIS transverse single spin asymmetry (SSA) 
$A^{\sin(\phi_h-\phi_S)}_{UT}$ measured by HERMES and COMPASS,
in the $\gamma^* - p$ c.m. frame and at order $k_\perp/Q$, is given 
by~\cite{Anselmino:2005ea, Anselmino:2005nn}:
\be
A^{\sin (\phi_h-\phi_S)}_{UT} = 
\label{hermesut}
\frac{\displaystyle  \sum_q \int
{d\phi_S \, d\phi_h \, d^2 \bfk _\perp}\;
\Delta^N \! f_{q/\pup} (x,\kt) \sin (\varphi -\phi_S) \;
\frac{d \hat\sigma ^{\ell q\to \ell q}}{dQ^2} \;
\; D_q^h(z,p_\perp) \sin (\phi_h -\phi_S) }
{\displaystyle \sum_q \int {d\phi_S \,d\phi_h \, d^2 \bfk _\perp}\;
f_{q/p}(x,k _\perp) \; \frac{d \hat\sigma ^{\ell q\to \ell q}}{dQ^2} \;
 \; D_q^h(z,p_\perp) } \> \cdot
\ee
$\phi_S$ and $\phi_h$ are the azimuthal angles identifying the
directions of the proton spin $\bfS$ and of the momentum of the outgoing hadron $h$
respectively, while $\varphi$ defines the direction of the incoming
(and outgoing) quark transverse momentum,
$\bfk_\perp$ = $\kt(\cos\varphi, \sin\varphi,0)$;
$\frac{d \hat\sigma ^{\ell q\to \ell q}}{dQ^2}$ is the unpolarized
cross section for the elementary scattering  $\ell q\to \ell q$.
%
%
%
%
\begin{figure}[t]
\vspace*{-2pc} 
\hspace{3.5pc}
\includegraphics[width=13pc]{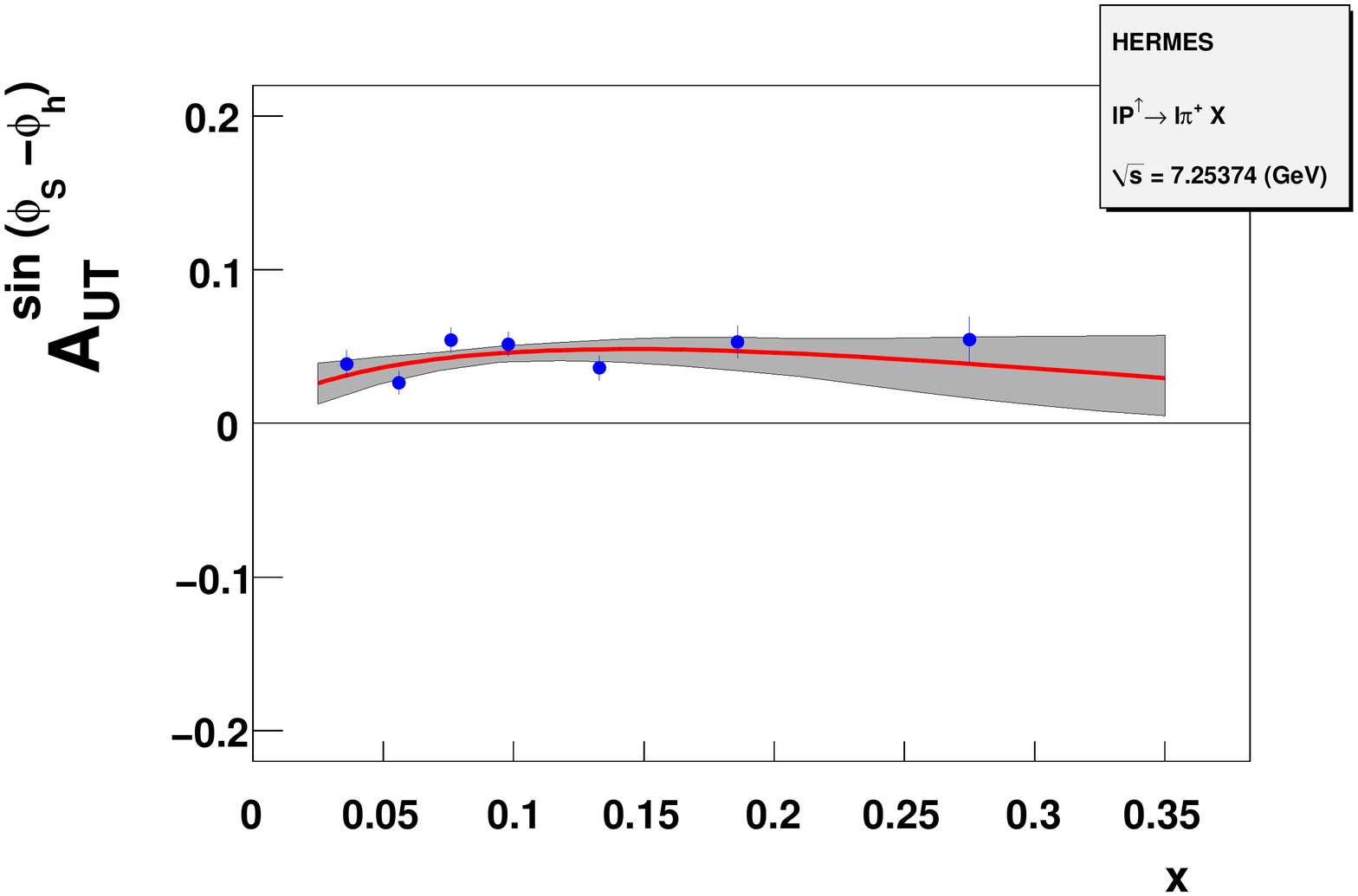}
\hspace{3pc}
\includegraphics[width=13pc]{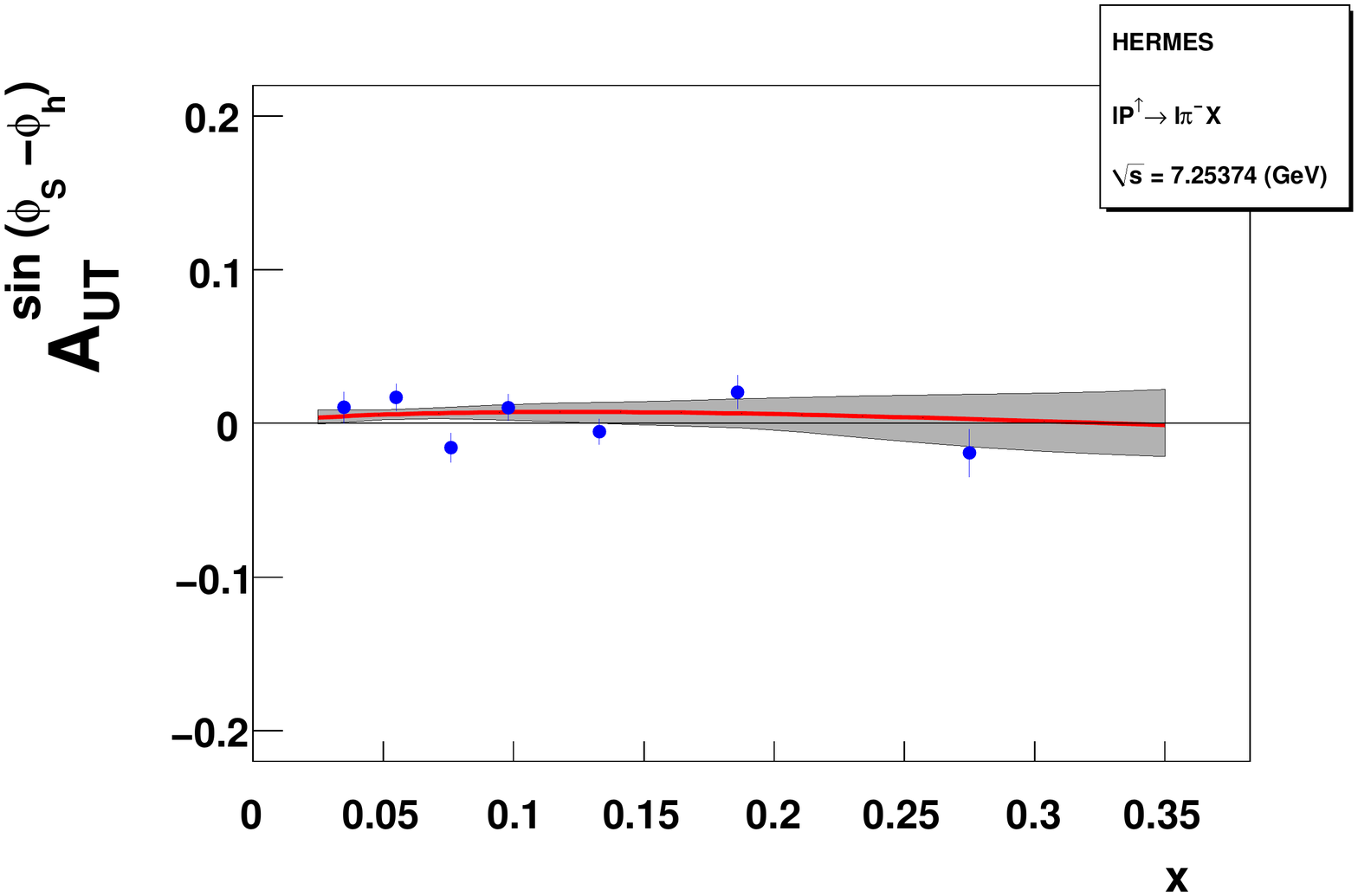}
\vspace*{-2pc} 
\\  
\hspace*{3.5pc} 
\includegraphics[width=13pc]{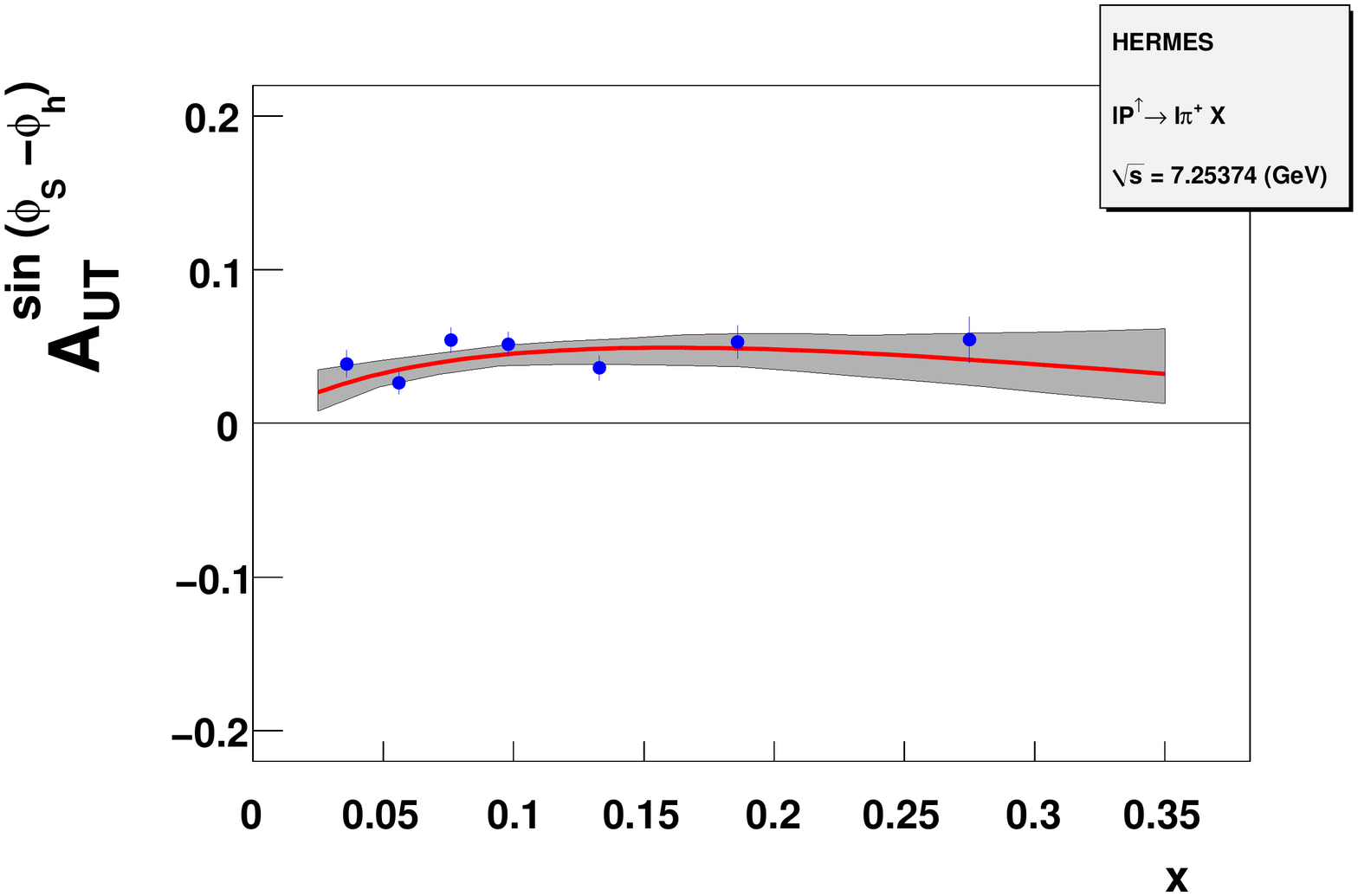}
\hspace{3pc}
\includegraphics[width=13pc]{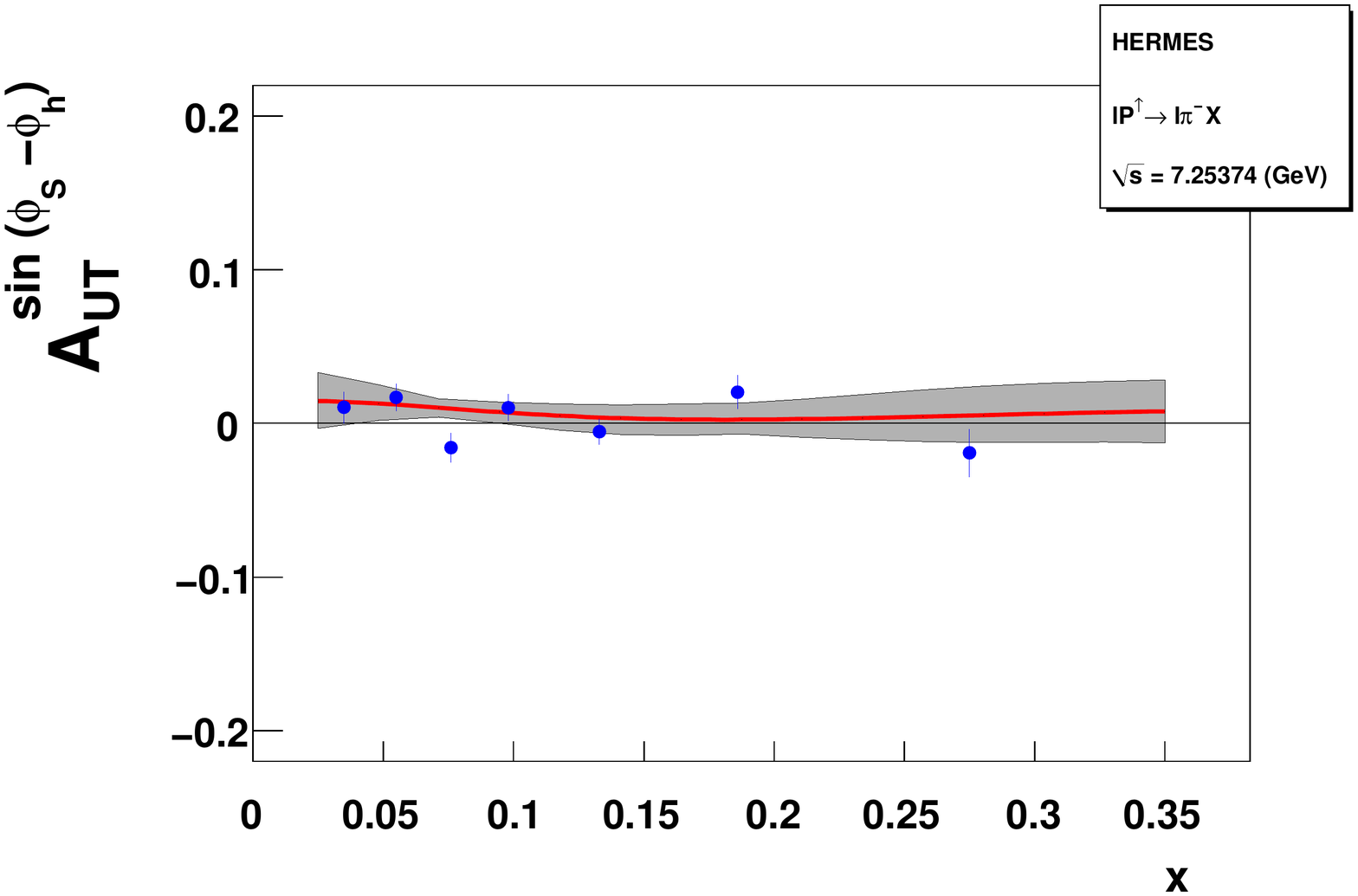}
\vspace*{-1pc} 
\caption{\label{herm-pi}The Sivers single spin asymmetry, $A_{UT}^{\sin(\phi_h-\phi_S)}$, as a function of $x$ for SIDIS production of $\pi^+$ (left panels) and $\pi^-$ (right panels) at HERMES. The upper (lower) panels show the results obtained from a fit which includes only valence (valence and sea) quark contributions.}
\end{figure}
%
%
Finally, $D_q^h(z,p_\perp)$ is the fragmentation function describing the
hadronization of the final quark $q$ into the detected hadron $h$ with
momentum $\bfP_h$; $h$ carries, with respect
to the fragmenting quark, a light-cone momentum fraction $z$ and a transverse
momentum $\bfp_\perp$.

The Sivers function is parameterized in terms of the unpolarized
distribution function, as in Ref.~\cite{Anselmino:2005ea}, in the following
factorized form:
\be
\Delta^N \! f_ {q/\pup}(x,\kt) = 2 \, {\cal N}_q(x) \, h(\kt) \,
f_ {q/p} (x,\kt)\; , \label{sivfac}
\ee
with
\be
{\cal N}_q(x) =  N_q \, x^{\alpha_q}(1-x)^{\beta_q} \,
\frac{(\alpha_q+\beta_q)^{(\alpha_q+\beta_q)}}
{\alpha_q^{\alpha_q} \beta_q^{\beta_q}}\; ,
\hspace{2pc} {\rm and} \hspace{2pc} 
h(\kt) = \sqrt{2e}\,\frac{k_\perp}{M_{1}}\,e^{-{k_\perp^2}/{M_{1}^2}}\; ,
\ee
where $N_q$, $\alpha_q$, $\beta_q$ and $M_1$ (GeV/$c$) are free parameters
to be determined by fitting the experimental data. 
We adopt a Gaussian factorization for the unpolarized distribution and 
fragmentation functions with the Gaussian widths $\langle k_\perp^2\rangle$ 
and $\langle p_\perp^2\rangle$ fixed to the values found in 
Ref.~\cite{Anselmino:2005nn} by analysing the Cahn effect in unpolarized 
SIDIS: $\langle\kt^2\rangle = 0.25 \;({\rm GeV}/c)^2$ and 
$\langle p_\perp^2\rangle  = 0.20 \;({\rm GeV}/c)^2$. For the unpolarized, $\kt$-integrated distribution and fragmentation functions we use the GRV98~\cite{Gluck:1998xa} and DSS~\cite{deFlorian:2007aj} sets.

We best fit the HERMES proton and COMPASS deuteron data from 
Refs.~\cite{:2008dn,:2009ti} and exclude the COMPASS proton 
data~\cite{Alekseev:2010rw} due to the presence of some experimental scale uncertainty which is difficult to account for in the present analysis. The new forthcoming data from COMPASS on a proton target will become very useful in the future.

\begin{figure}[t]
\vspace*{-2pc} 
\hspace*{3.5pc}
\includegraphics[width=13pc]{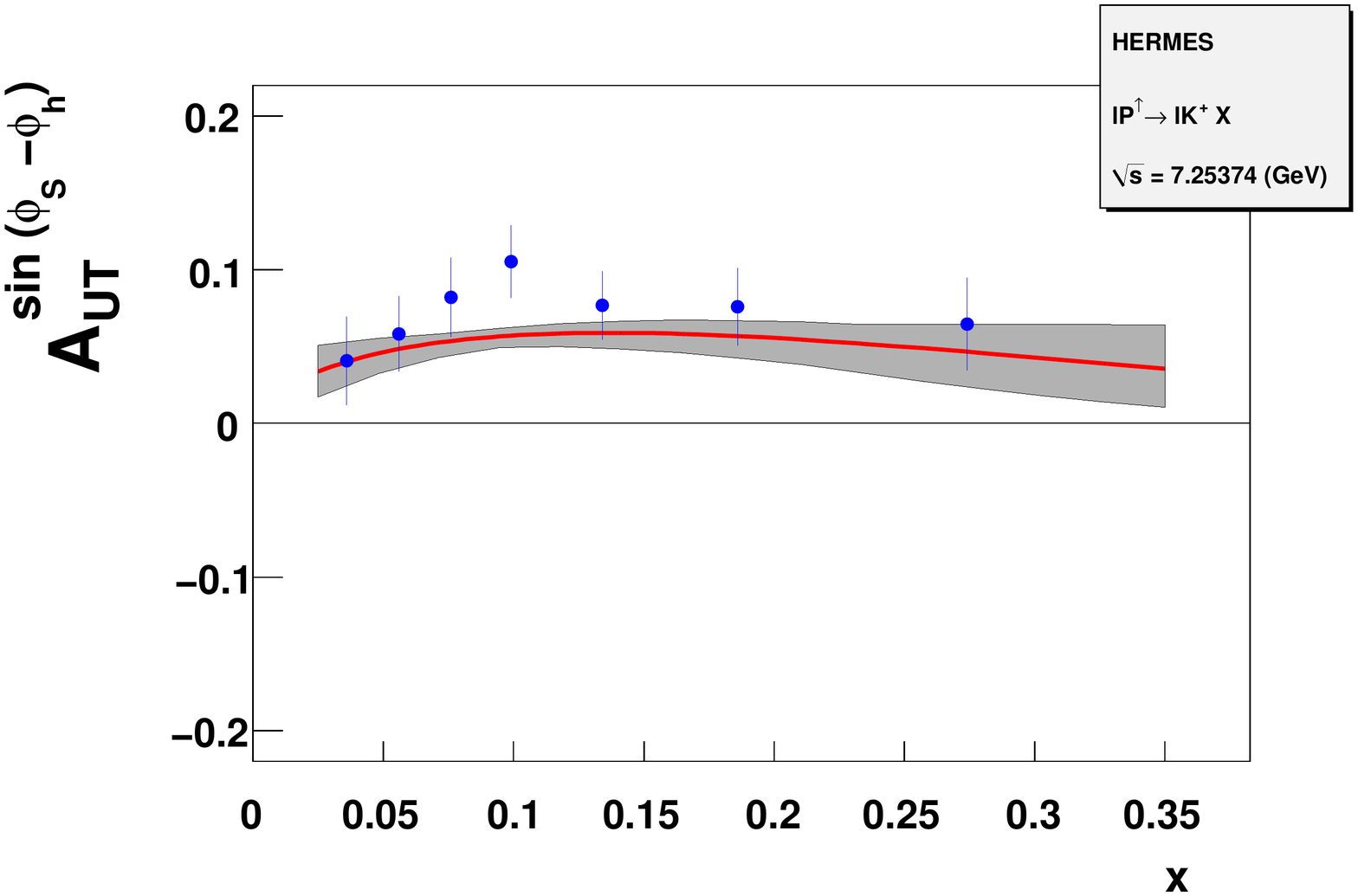}
\hspace{3pc}
\includegraphics[width=13pc]{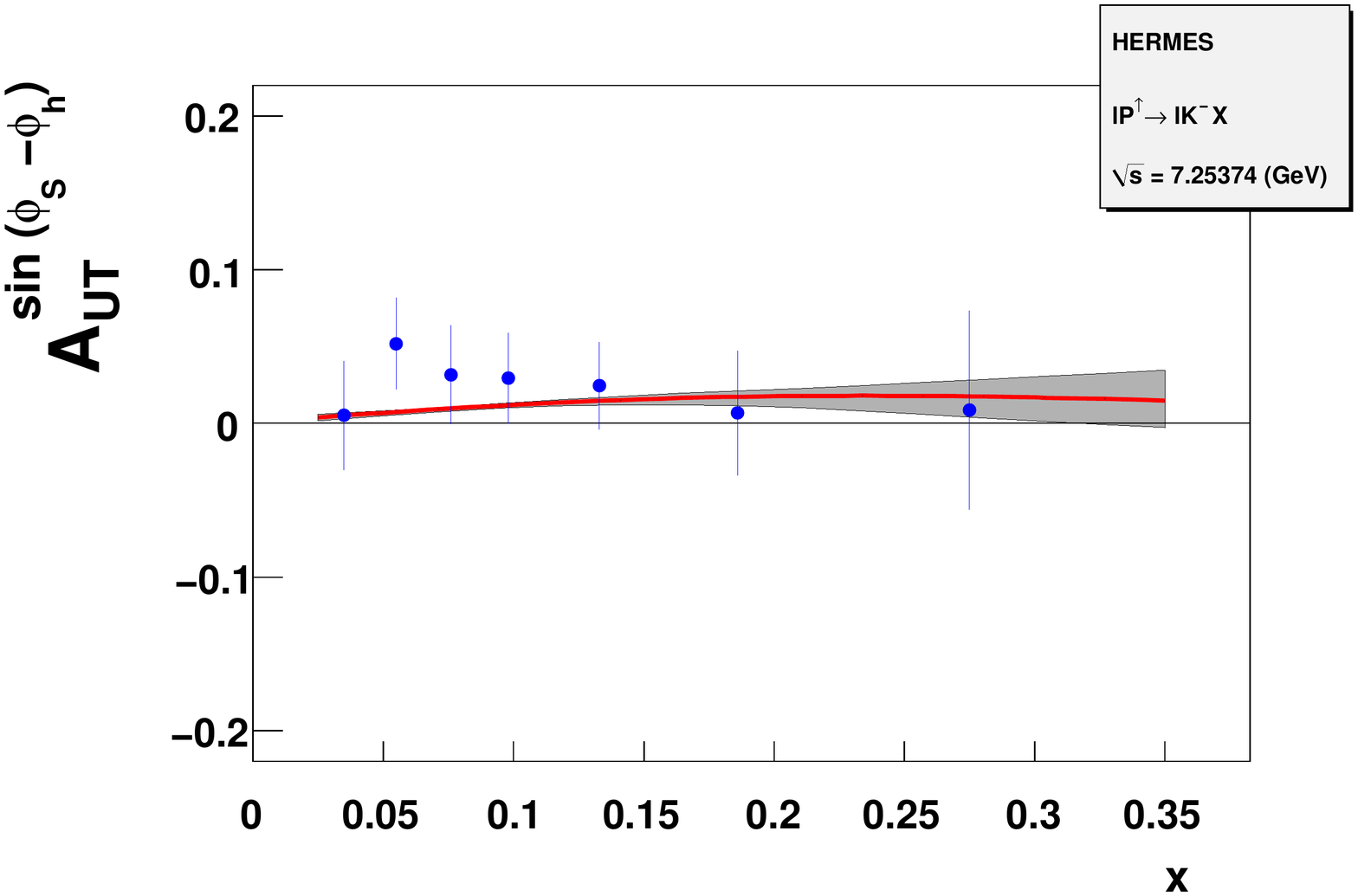}
\vspace*{-2pc} 
\\
\hspace*{3.5pc}
\includegraphics[width=13pc]{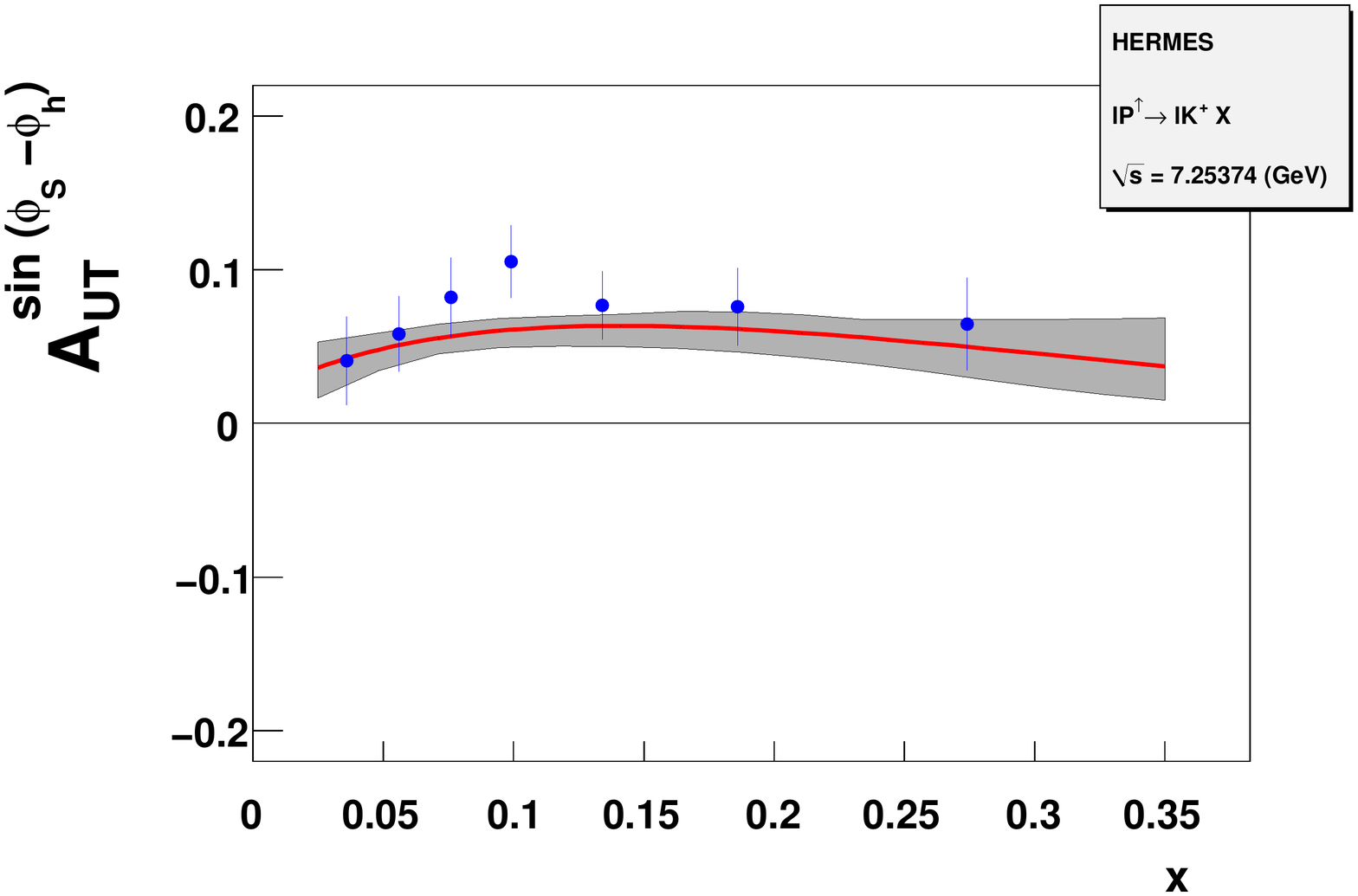}
\hspace{3pc}
\includegraphics[width=13pc]{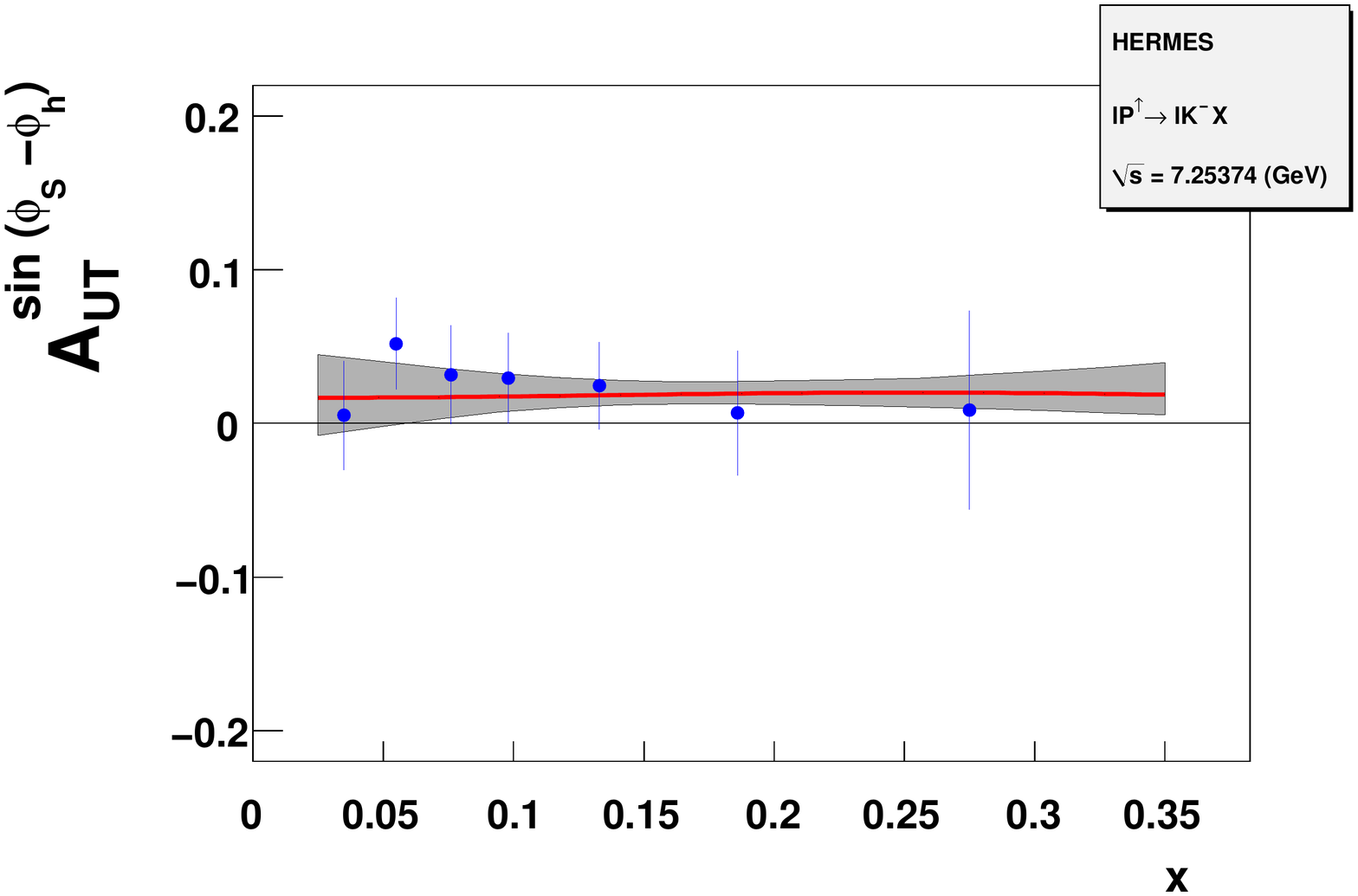}
\vspace*{-1pc} 
\caption{\label{herm-K} The Sivers single spin asymmetry, $A_{UT}^{\sin(\phi_h-\phi_S)}$, as a function of $x$ for SIDIS production of $K^+$ (left panels) and $K^-$ (right panels) at HERMES. The upper (lower) panels show the 
results obtained from a fit which includes only valence (valence plus sea)
quark contributions.}
\end{figure}

\begin{figure}[t]
\vspace*{-2pc} 
\hspace*{3.5pc}
\includegraphics[width=13pc]{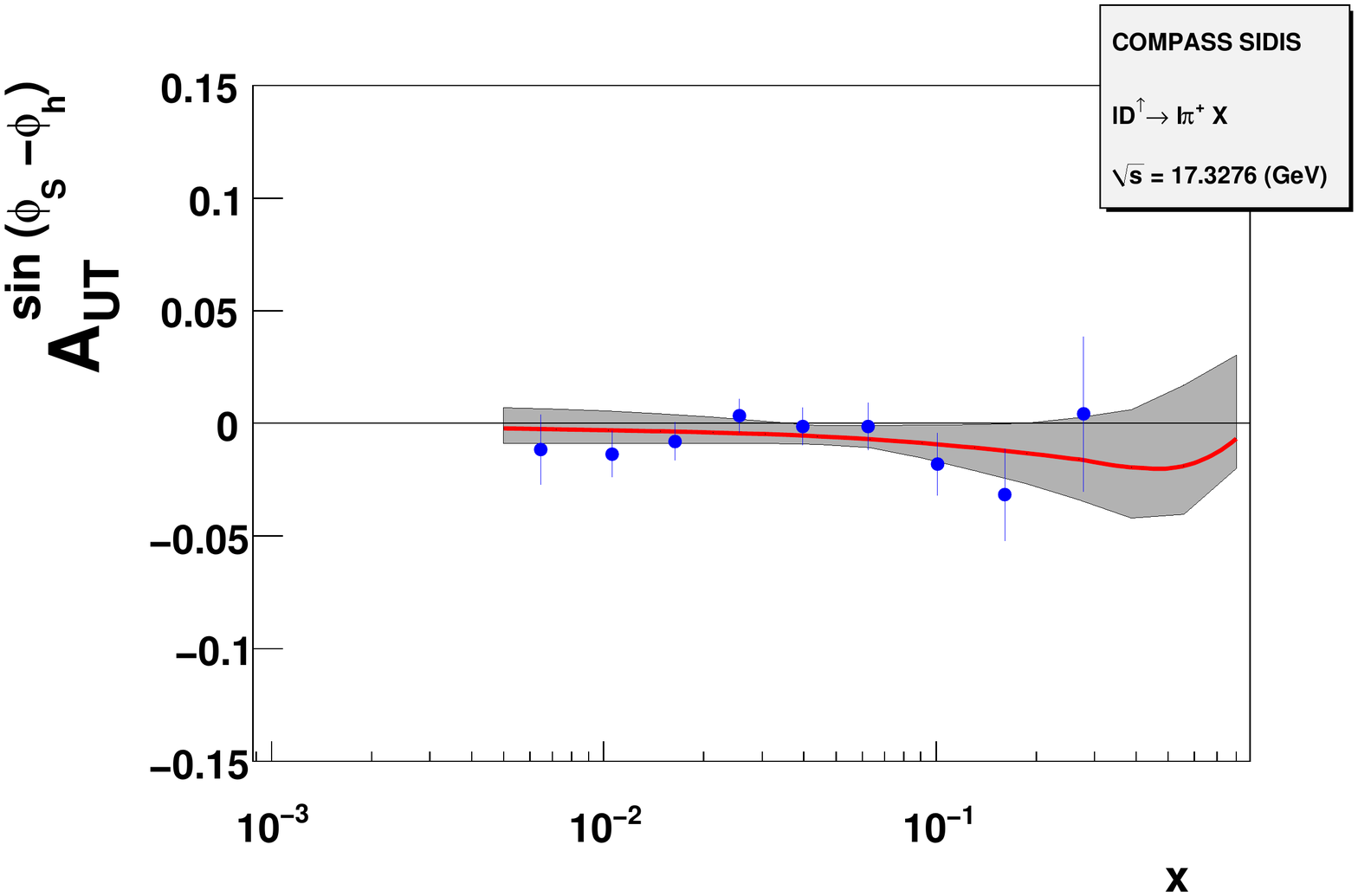}
\hspace{3pc}
\includegraphics[width=13pc]{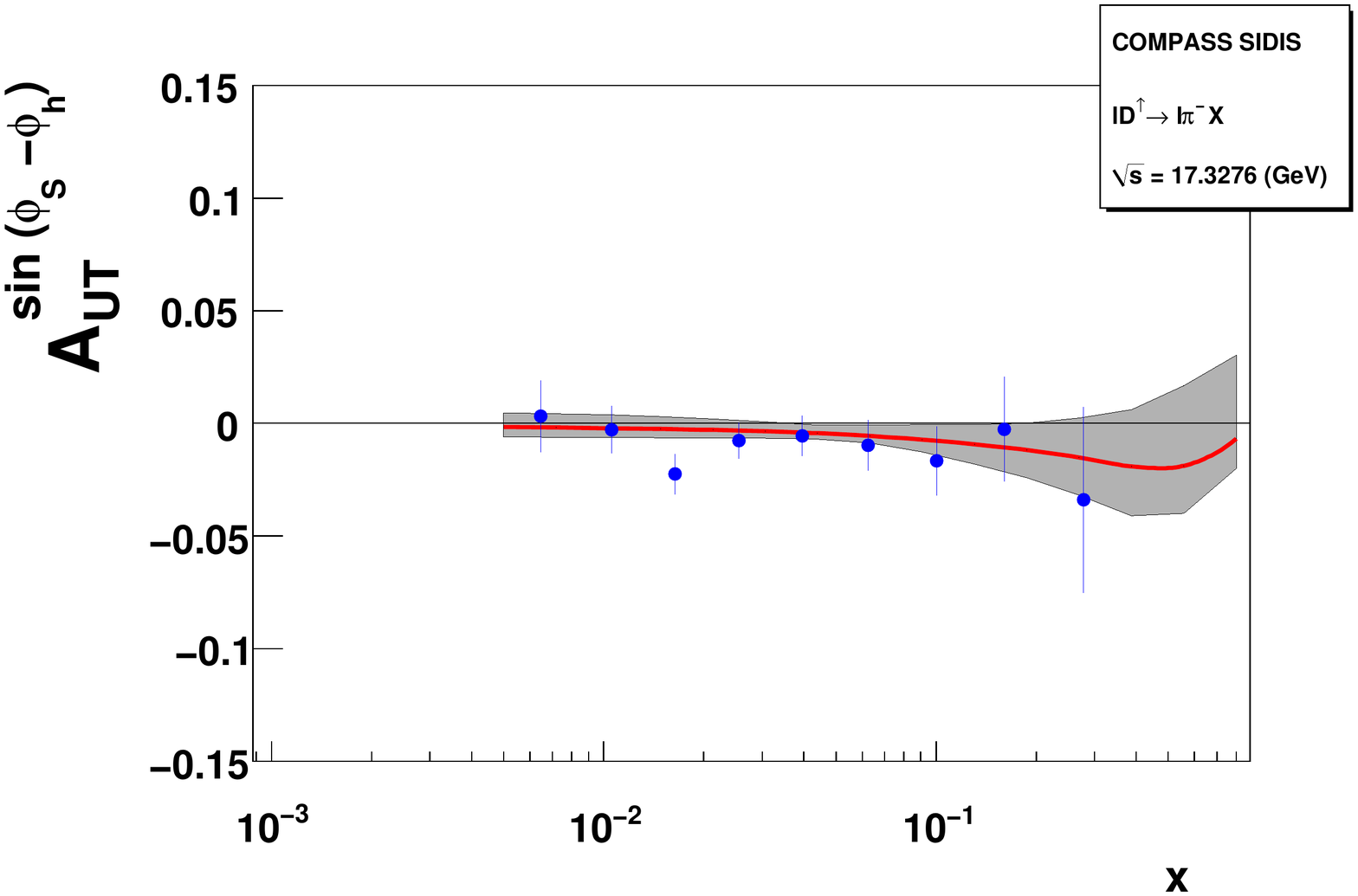}
\vspace*{-2pc} 
\\
\hspace*{3.5pc}
\includegraphics[width=13pc]{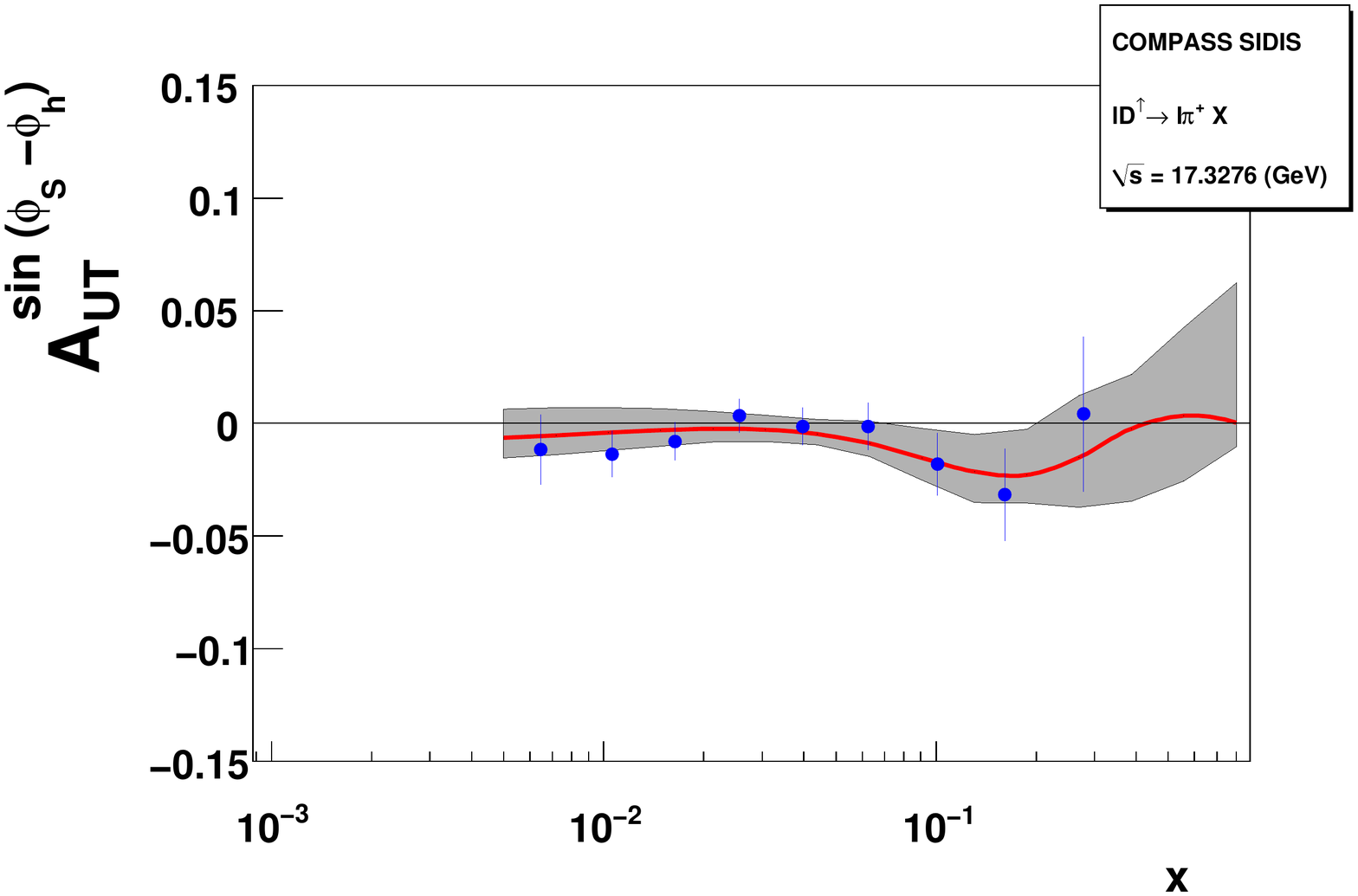}
\hspace{3pc}
\includegraphics[width=13pc]{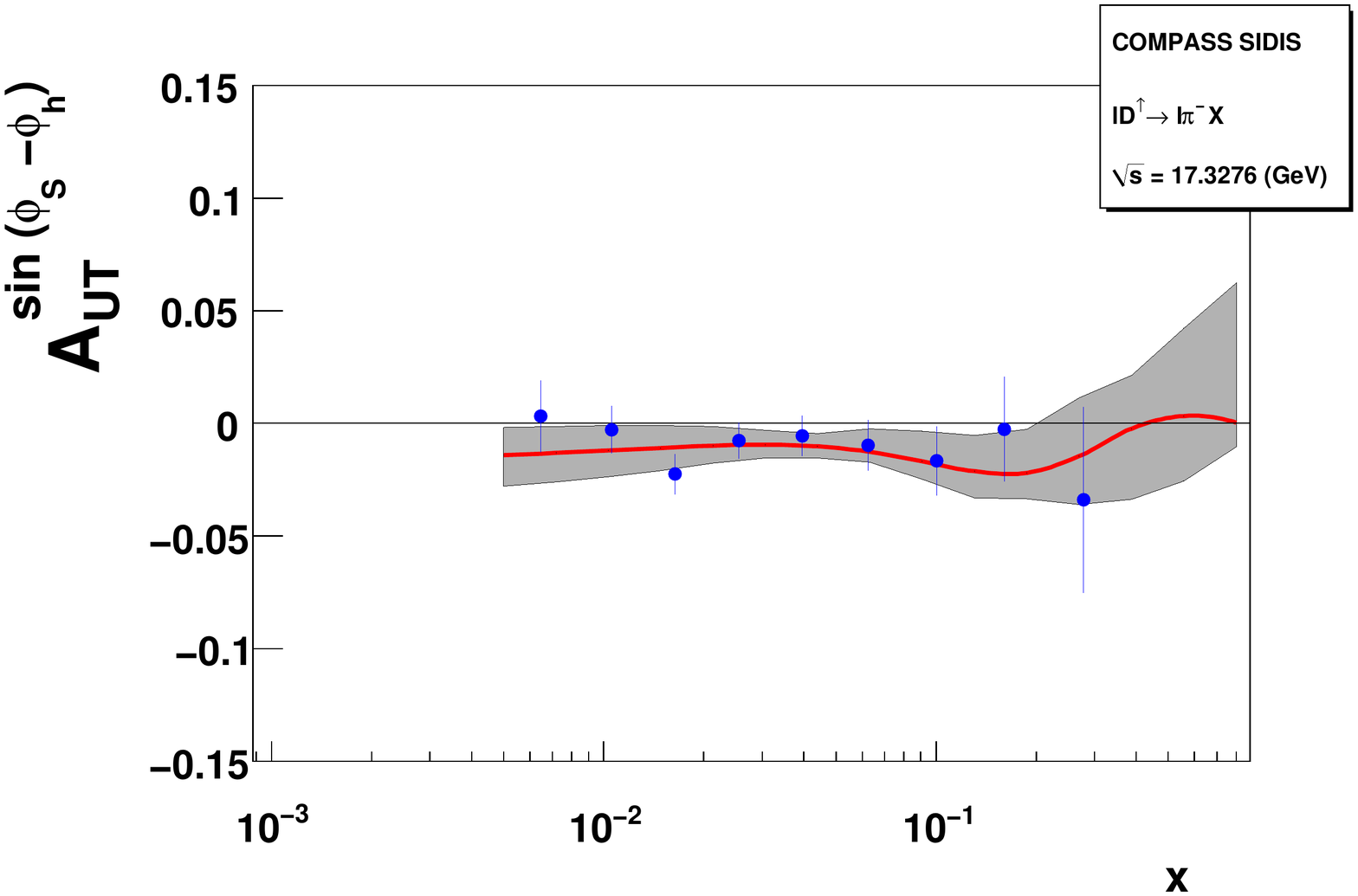}
\vspace*{-1pc} 
\caption{\label{comp-pi}The Sivers single spin asymmetry $A_{UT}^{\sin(\phi_h-\phi_S)}$, as a function of $x$ for SIDIS 
$\pi^+$  (left panels) and $\pi^-$ (right panels) production at COMPASS on a deuteron target. The upper (lower) panels show the results obtained from a fit which includes only valence (valence plus sea) quark contributions.}
\end{figure}

\begin{figure}[b]
\vspace*{-1pc} 
\hspace*{3.5pc}
\includegraphics[width=13pc]{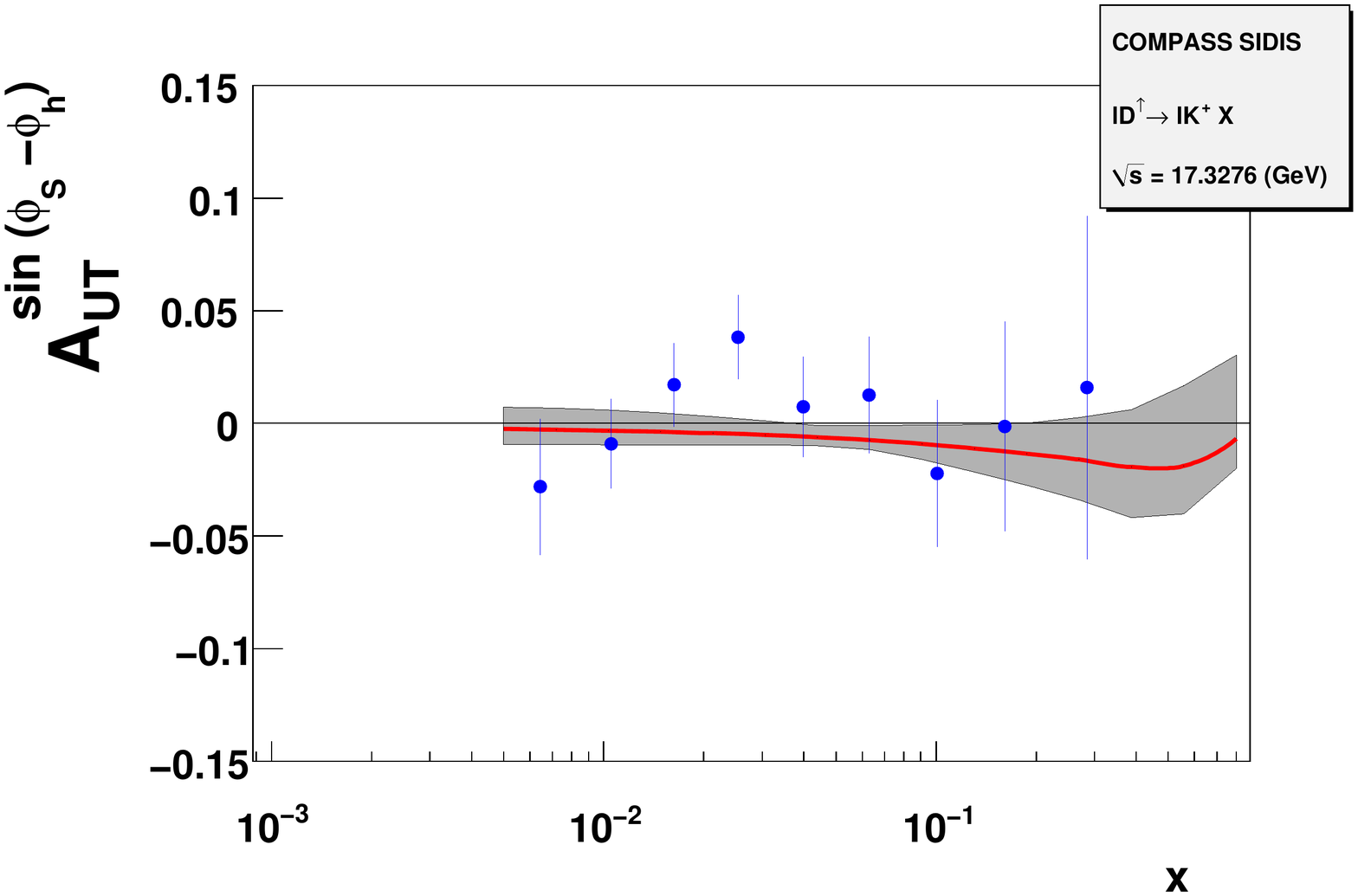}
\hspace{3.5pc}
\includegraphics[width=13pc]{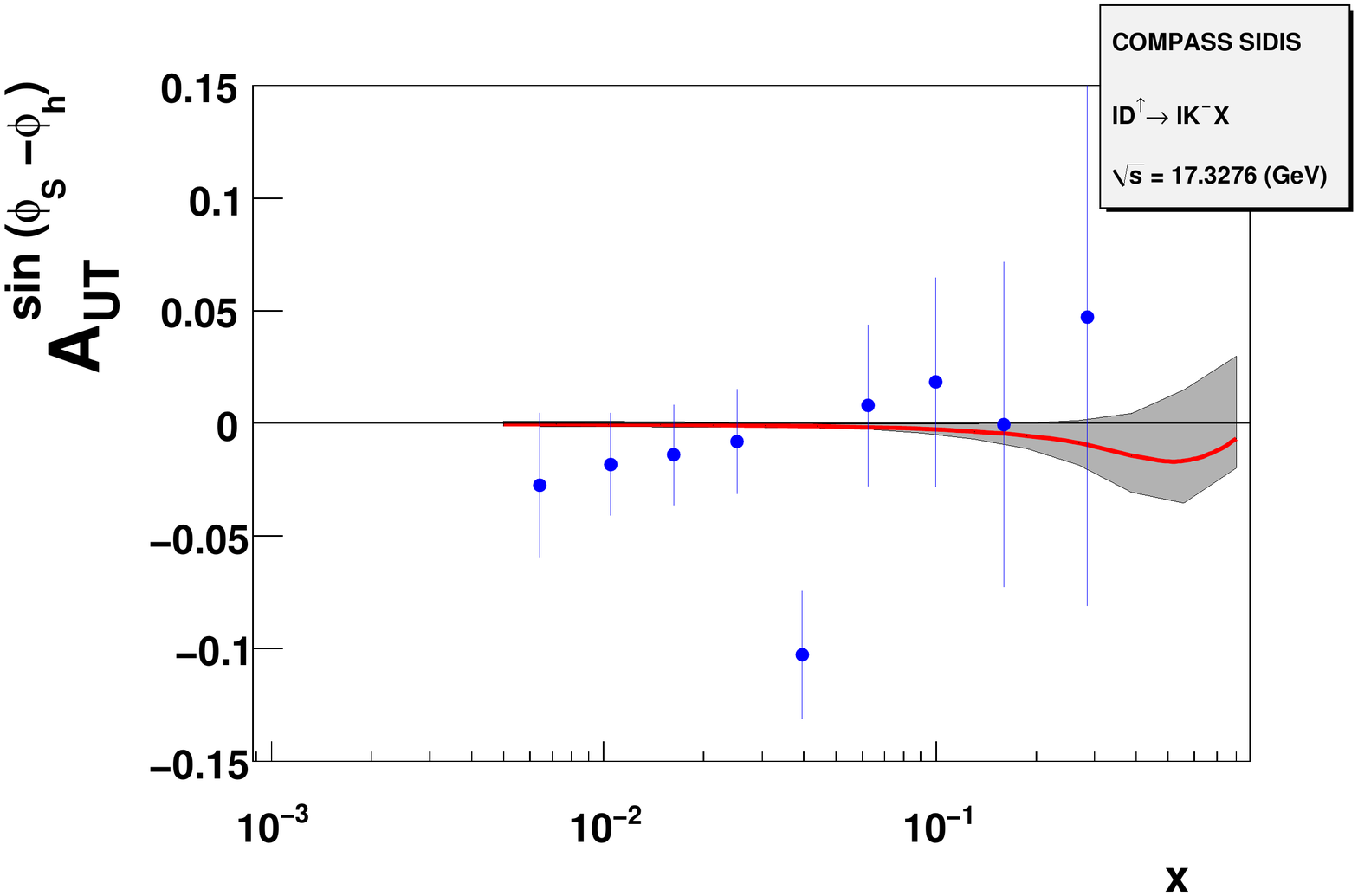}
\vspace*{-2pc} 
\\
\hspace*{3.5pc}
\includegraphics[width=13pc]{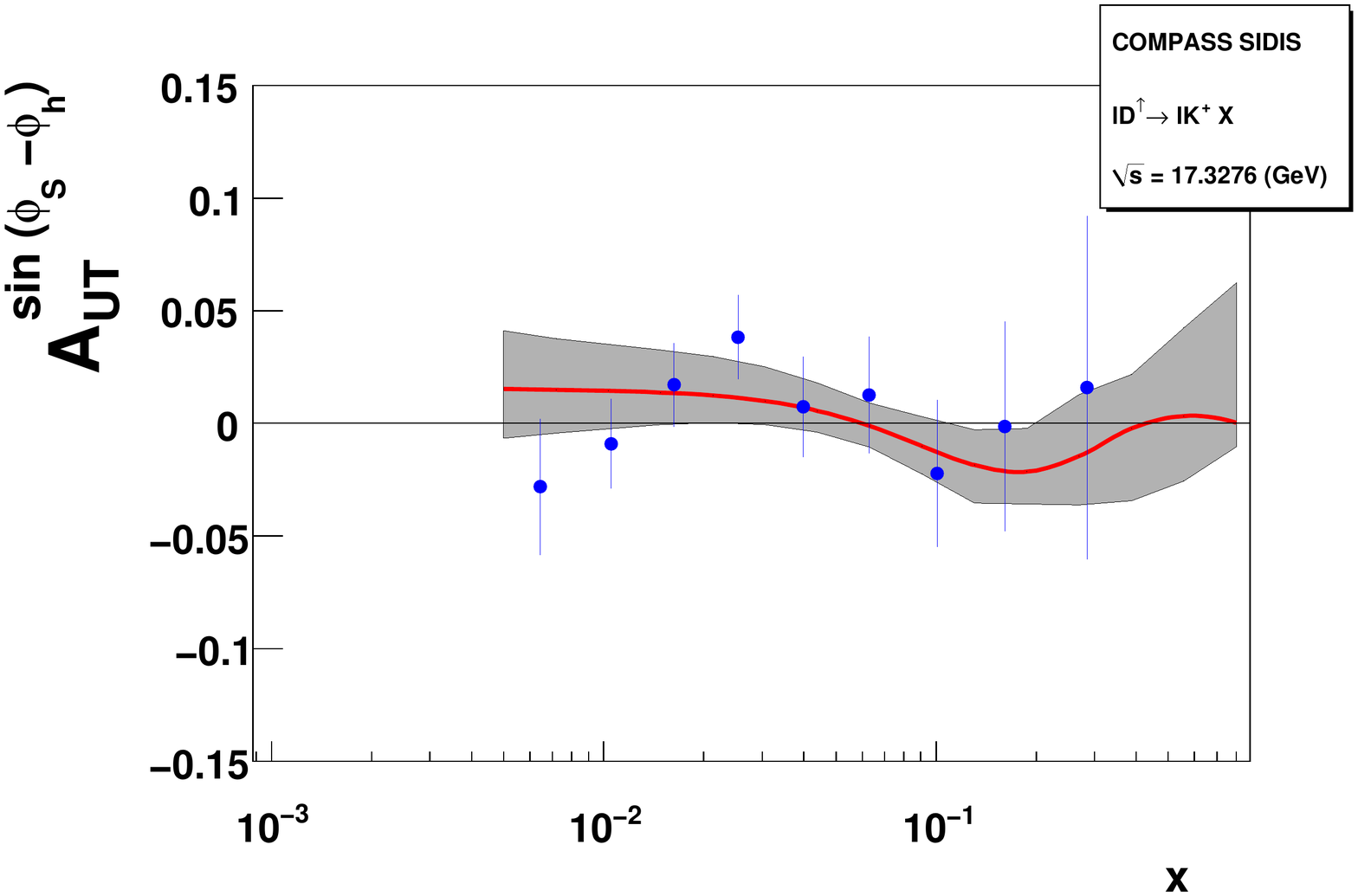}
\hspace{3.5pc}
\includegraphics[width=13pc]{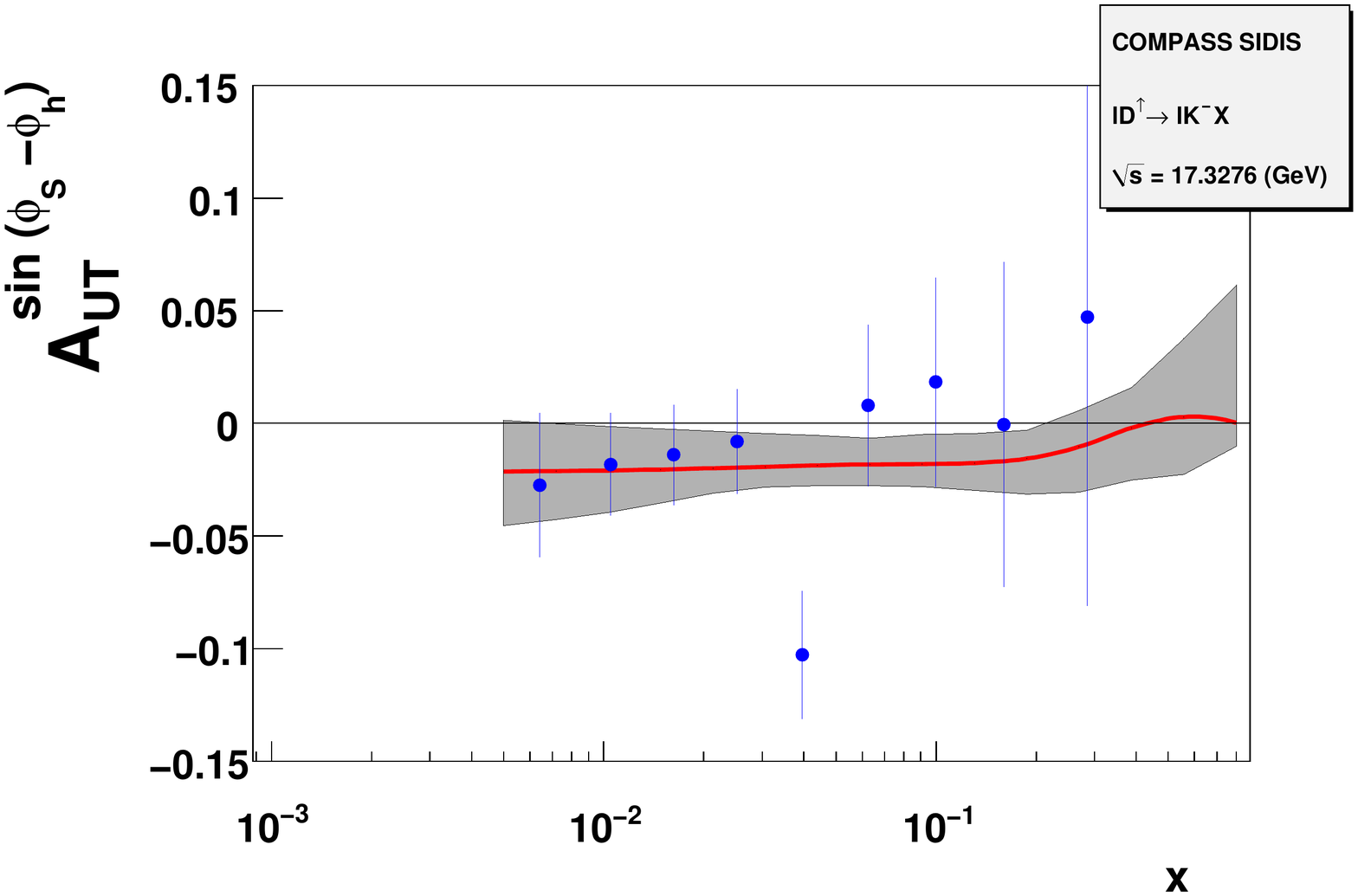}
\vspace*{-1pc} 
\caption{\label{comp-K} The Sivers single spin asymmetry $A_{UT}^{\sin(\phi_h-\phi_S)}$, as a function of $x$ for SIDIS $K^+$ (left panels) and $K^-$ (right panels) production at COMPASS on a deuteron target. The upper (lower) panels show the results obtained from a fit which includes only valence (valence plus sea) quark contributions.}
\end{figure}

\begin{figure}[b]
\vspace{2pc} 
\hspace{3pc}
\includegraphics[width=13pc]{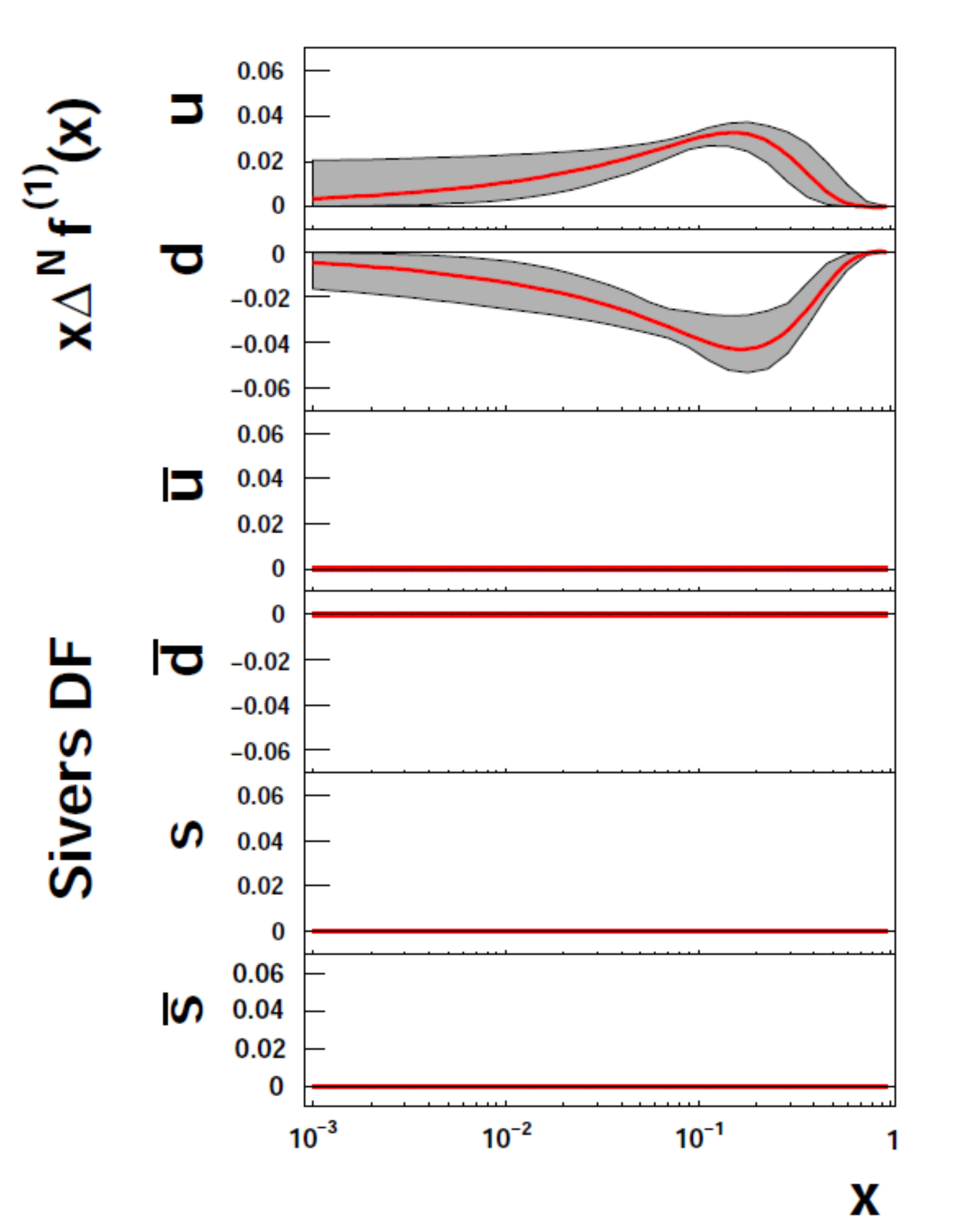}
\hspace{3.3pc}
\includegraphics[width=13pc]{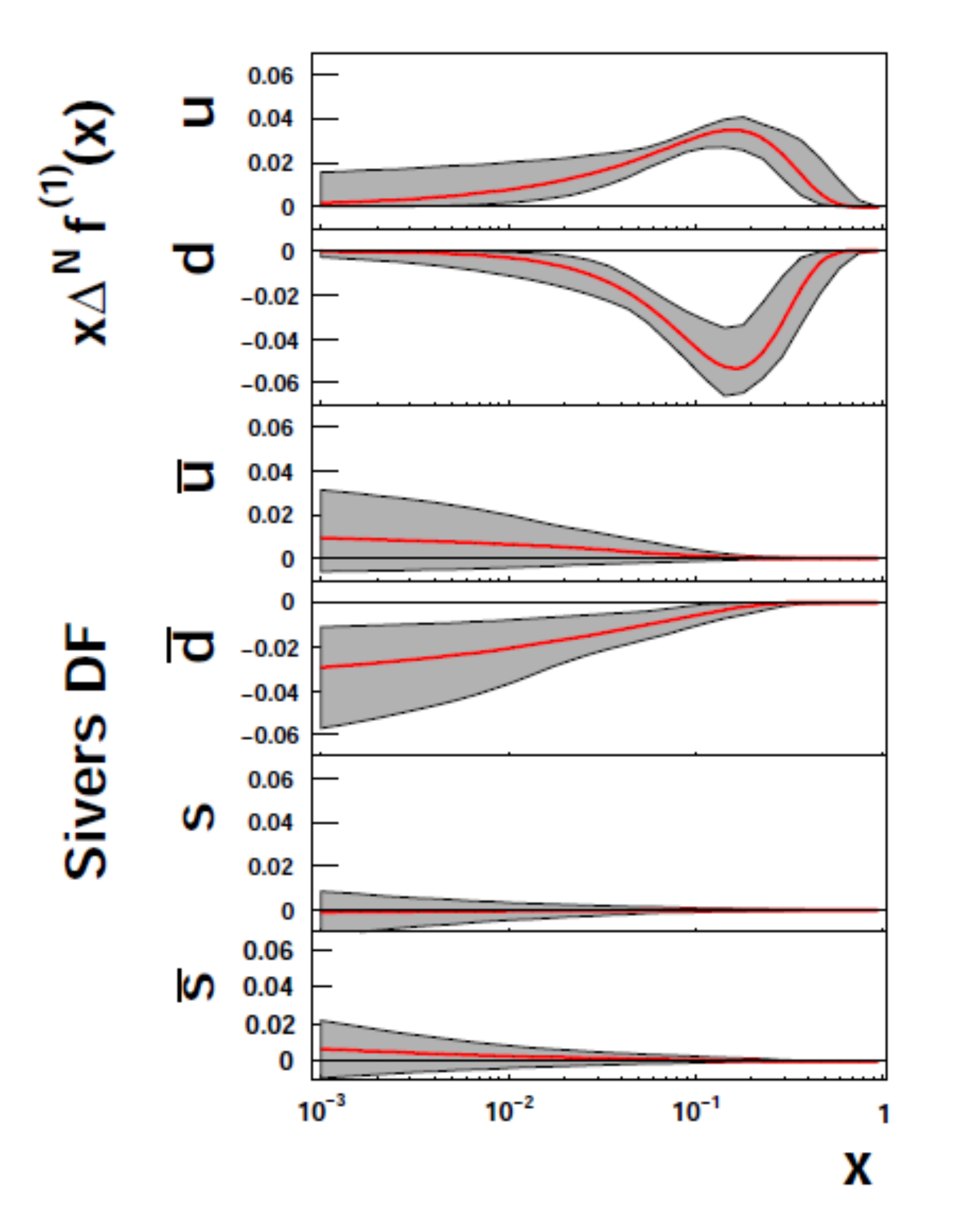}
\vspace*{-0.5pc} 
\caption{\label{Sivers-fn} First $\kt$-moment of the Sivers distribution function,  $x\Delta^N f^{(1)}_{q/p^\uparrow}$, as a function of $x$ as 
obtained by fitting SIDIS experimental data from HERMES and COMPASS. The left (right) panel shows the results obtained from a fit  which includes only valence (valence plus sea) quark contributions.}
\end{figure}

In order to evaluate the significance of the sea-quark Sivers contributions 
we first perform a fit of the SIDIS data including only Sivers functions for $u$ and $d$ quarks. The results we obtain are rather satisfactory, with a 
total $\chi^2_{dof}$ of about $1.05$. They are shown in the upper panels 
of Figs.~\ref{herm-pi} -- \ref{comp-K}. The corresponding Sivers functions 
are plotted in the left panel of Fig.~\ref{Sivers-fn}. It is interesting to 
notice that the $M_1$ parameter, which fixes the Gaussian width of the 
Sivers function (i.e. its distribution in $\kt$) is rather well constrained and turns out 
to be between one half and two thirds of the unpolarized distribution 
function width.
%
%
%

We then perform a second fit, in which we allow for $\bar u$, $\bar d$,  
$s$, $\bar s$ sea contributions to the Sivers function: overall, we obtain 
similar results, shown in the lower panels of Figs.~\ref{herm-pi} -- \ref{comp-K}, but the total $\chi^2_{dof}$ improves to about $0.9$. 
The corresponding Sivers functions are plotted in the right panel of 
Fig.~\ref{Sivers-fn}.

At this stage, our results can only be considered as qualitative and preliminary, as the experimental data are not yet so well established 
and abundant to allow a precise definite determination of the Sivers sea distribution functions. We can only try, by comparing the upper and lower 
panels of Figs.~\ref{herm-pi} -- \ref{comp-K}, to draw some general 
conclusions. 
%
%
%
%
%
%

Adding the sea quark contributions to the Sivers functions does not induce 
any variation in the $A^{\sin (\phi_h-\phi_S)}_{UT}$ azimuthal moments 
for $\pi^+$ and $K^+$ production, which are confirmed to be essentially 
generated by the fragmentation of a valence $u$ quark from the proton; 
instead, the production of $\pi^-$ and mostly $K^-$ are more sensitive 
to the presence of the sea quark contributions, and might offer the chance 
to gain some preliminary information on the polarized proton sea.
 
We also learn that the $A^{\sin (\phi_h-\phi_S)}_{UT}$ asymmetries, as delivered by the presently available experimental data, are not very 
sensitive to the strange quark Sivers distributions, while they feel the 
presence of $\bar u$ and $\bar d$ contributions more strongly: 
we have actually checked that the $\chi^2_{dof}$ is not altered when 
adding or removing  $s$ and $\overline s$.  

When comparing the left and right plots in Fig.~\ref{Sivers-fn}, one 
notices that, while the $\bar u$, $s$ and $\bar s$ Sivers functions are consistent with zero, the $\bar d$ Sivers function can be rather large and 
negative, definitely not consistent with zero. This originates from the 
small-$x$ behavior of the COMPASS data for $\pi^+$ production off a 
deuteron target, most sensitive to $\bar d$: as these data show a negative 
trend at small $x$ (and very tiny error bars, Fig.~\ref{comp-pi}), the fit 
naturally tends to assign a negative value to the $\bar d$ Sivers function.

Further experimental measurements from present and future facilities are 
strongly needed to reach a better knowledge on the important Sivers 
distribution functions, in particular in the large and small $x$ regions 
where they are presently largely unconstrained. Some help might also come 
from the existing large SSAs in $p \, p \to h \, X$ processes, which can be 
interpreted within a TMD factorization scheme 
\cite{D'Alesio:2004up, Anselmino:2005sh}.

%
%
%
\section*{References}
\bibliography{sample}

\providecommand{\newblock}{}
\begin{thebibliography}{10}
\expandafter\ifx\csname url\endcsname\relax
  \def\url#1{{\tt #1}}\fi
\expandafter\ifx\csname urlprefix\endcsname\relax\def\urlprefix{URL }\fi
\providecommand{\eprint}[2][]{\url{#2}}

\bibitem{Anselmino:2008sga}
Anselmino M {\em et~al.\/} 2009 {\em Eur. Phys. J.\/} {\bf A39} 89--100
  (\textit{Preprint} \eprint{0805.2677})

\bibitem{Diefenthaler:2007rj}
Diefenthaler M (HERMES) 2007  (\textit{Preprint} \eprint{arXiv:0706.2242
  [hep-ex]})

\bibitem{:2008dn}
Alekseev M {\em et~al.\/} (COMPASS) 2009 {\em Phys. Lett.\/} {\bf B673}
  127--135 (\textit{Preprint} \eprint{0802.2160})

\bibitem{Alekseev:2010rw}
Alekseev M~G {\em et~al.\/} (The COMPASS) 2010 {\em Phys. Lett.\/} {\bf B692}
  240--246 (\textit{Preprint} \eprint{1005.5609})

\bibitem{:2009ti}
Airapetian A {\em et~al.\/} (HERMES) 2009 {\em Phys. Rev. Lett.\/} {\bf 103}
  152002 (\textit{Preprint} \eprint{0906.3918})

\bibitem{Anselmino:2005ea}
Anselmino M {\em et~al.\/} 2005 {\em Phys. Rev.\/} {\bf D72} 094007
  (\textit{Preprint} \eprint{hep-ph/0507181})

\bibitem{Anselmino:2005nn}
Anselmino M {\em et~al.\/} 2005 {\em Phys. Rev.\/} {\bf D71} 074006
  (\textit{Preprint} \eprint{hep-ph/0501196})

\bibitem{Gluck:1998xa}
Gluck M, Reya E and Vogt A 1998 {\em Eur. Phys. J.\/} {\bf C5} 461--470
  (\textit{Preprint} \eprint{hep-ph/9806404})

\bibitem{deFlorian:2007aj}
de~Florian D, Sassot R and Stratmann M 2007 {\em Phys. Rev.\/} {\bf D75} 114010
  (\textit{Preprint} \eprint{hep-ph/0703242})

\bibitem{D'Alesio:2004up}
D'Alesio U and Murgia F 2004 {\em Phys. Rev.\/} {\bf D70} 074009
  (\textit{Preprint} \eprint{hep-ph/0408092})

\bibitem{Anselmino:2005sh}
Anselmino M {\em et~al.\/} 2006 {\em Phys. Rev.\/} {\bf D73} 014020
  (\textit{Preprint} \eprint{hep-ph/0509035})

\end{thebibliography}
%
%
%
%

%
%
%

\end{document}